\documentclass[a4paper]{article}

\usepackage{amsmath,amsfonts,amssymb}
\usepackage{amsthm}
\usepackage{enumerate}
\usepackage{bbm}
\usepackage{graphicx}
\usepackage{lscape,epsf}
\usepackage[vcentermath]{youngtab}
\usepackage{rotating}
\usepackage{hyperref}

\AtBeginDocument{\addtolength{\columnsep}{8pt}}
\AtBeginDocument{\addtolength{\textwidth}{10pt}}
\def\ds{\displaystyle}
\def\bea{\begin{array}{c}}
\def\ea{\end{array}}
\def\be{\begin{equation}\bea\ds}
\def\ee{\ea\end{equation}}
\def\bee{\begin{equation}\begin{array}{rcl}\ds}
\def\eee{\end{array}\end{equation}}

\def\nb{{\bf n}}
\def\bb{{\bf b}}
\def\tb{{\bf t}}

\def\tr{{\rm Tr}\,}
\def\nn{\nonumber}

\def\htau{\hat{\tau}}
\def\hk{\hat{\kappa}}

\def\eb[#1]{{\bf e}_{#1}}
\def\Ab{{\bf A}}

\newcommand\Trule{\rule{0pt}{2.6ex}}
\newcommand\Brule{\rule[-1.2ex]{0pt}{0pt}}

\title{\bf Chern-Simons  Improved Hamiltonians \\
            for Strings in Three Space Dimensions}

\author{Ivan Gordeli$^{a}$, Dmitry Melnikov$^{a,b}$, Antti Niemi$^{c,d,e}$ and Ara Sedrakyan$^{f,g}$}

\date{}

\begin{document}



\twocolumn[
  \begin{@twocolumnfalse}
  \hfill{ITEP-TH-18/15}
    \maketitle
    \vspace{-0.5cm}
    \begin{center}
    {\textit{\small $^a$  International Institute of Physics, Federal University of Rio Grande do Norte, \\Av. Odilon Gomes de Lima 1722, Capim Macio, Natal-RN  59078-400, Brazil}\\ \vspace{6pt}
\textit{\small $^b$  Institute for Theoretical and Experimental Physics, \\B.~Cheremushkinskaya 25, Moscow 117218, Russia}\\ \vspace{6pt}
\textit{\small $^c$  Department of Physics and Astronomy, Uppsala University,\\ P.O. Box 803, S-75108, Uppsala, Sweden}\\ \vspace{6pt}
\textit{\small $^d$  Laboratoire de Mathematiques et Physique Theorique CNRS UMR 6083, \\ F\'ed\'eration Denis Poisson, Universit\'e de Tours, Parc de Grandmont, F37200, Tours, France}\\ \vspace{6pt}
\textit{\small $^e$  Department of Physics, Beijing Institute of Technology, \\ Haidian District, Beijing 100081, P. R. China}\\ \vspace{6pt}
\textit{\small $^f$  Yerevan Physics Institute,\\ Alikhanian Br. str. 2, Yerevan 36, Armenia}\\ \vspace{6pt}
\textit{\small $^g$  The Niels Bohr Institute, Copenhagen University,\\ Blegdamsvej 17, DK-2100 Copenhagen, Denmark}\\ \vspace{6pt}}
    \end{center}
    \vspace{0.2cm}
    \begin{abstract}
The Frenet equation governs the extrinsic geometry of a string in three-dimensional ambient space in terms of the curvature and torsion, which are both scalar functions under string reparameterisations. The description engages a local SO(2) gauge symmetry, which emerges from the invariance of the extrinsic
string geometry under local frame rotations around the tangent vector. Here we inquire how to construct the most general SO(2) gauge invariant Hamiltonian of strings, in terms of the curvature and torsion. The construction instructs us to introduce a long-range (self-) interaction between strings, which is mediated by a three dimensional bulk gauge field with a Chern-Simons self-interaction.  The results support the proposal that fractional statistics should be prevalent in the case of three dimensional string-like configurations.
\end{abstract}
\vspace*{0.7cm}
  \end{@twocolumnfalse}
  ]

The geometry of a string which is embedded in three dimensional Euclidean space $\mathbb R^3$, is a classic subject in differential geometry~\cite{Frenet:1852,Spivak}. String-like objects are similarly pivotal  to several subfields of Physics. Examples range from fundamental, confining and cosmic strings in high energy physics and cosmology~\cite{GSW,fund,Rickles} to vortices,  superconductors and plasmas in condensed and soft-matter physics~\cite{volovik,babaev}, and all the way to polymers and biological macromolecules such as DNA, RNA and proteins~\cite{Grosberg,Danielsson:2009qm,Chernodub:2010xz,Hu:2011wg,Molkenthin:2011}. In particular, string-like structures in three space dimensions display intricate topology including knots~\cite{kauffman} and knotted solitons~\cite{fadde}. A set of closed strings in $\mathbb R^3$ may even proceed by leapfrogging~\cite{leap,paul}, in a manner that engages exotic exchange statistics~\cite{prl}.

The geometry of a string in $\mathbb R^3$ is governed by the Frenet equation~\cite{Frenet:1852,Spivak}. This equation relates the coordinate description $\mathbf x(s) \in \mathbb R^3$ of a string with length $L$ and with $s\in [0,L]$ the (arc length) parameter, to a description in terms of the extrinsic geometry in terms of curvature $\kappa$ and torsion $\tau$;  we note that both $\kappa(s)$ and $\tau(s)$ transform as scalars under local reparameterisations $s \to \tilde s(s)$ of the string.

At a regular point along the string, where the curvature $\kappa(s) \not=0 $,  we may always unambiguously introduce  the Frenet frame which consists of an orthonormal triplet of the tangent, bi-normal and normal vectors $\{\tb,\bb,\nb\}$,
\begin{eqnarray*}
\Trule\Brule \tb & = & \frac{d{\bf x}/ds}{||d{\bf x}/ds||}\,,\\
\Trule\Brule \bb & = & \tb\times\frac{d^2{\bf x}/d s^2}{||d^2{\bf x}/d s^2||}\,,\\
\Trule\Brule \nb & = & \bb\times \tb\,.
\end{eqnarray*}
The Frenet equation then transports the Frenet frame along the string, in terms of a transfer matrix $R_F$ which is determined by the local curvature and torsion
$$
\frac{d}{ds}\left(
\begin{array}{c}
\nb \\
\bb \\
\tb
\end{array}
\right)  \ = \
R_F
\left(
\begin{array}{c}
\nb \\
\bb \\
\tb
\end{array}
\right).
$$
Explicitly,
\[
R_{F} \ \equiv \  \left(
\begin{array}{ccc}
0 & \tau(s) & -\kappa(s) \\
-\tau(s) & 0 & 0 \\
\kappa(s) & 0 & 0
\end{array}
\right).
\]
While the Frenet frame can not be introduced at an inflection point, where $d^2{\bf x}/ds = 0$,  there are alternative frames such as the Bishop's frame that do exist even at an inflection point~\cite{Hu:2011wg,Carroll}. At any regular point along the string we may then relate the Frenet frame to another, generic orthonormal frame of the form $\{\tb,\eb[1],\eb[2]\}$ by a $\text{SO(2)}\in\text{SO(3)}$ rotation $U$ around the vector $\tb(s)$ which is tangent to the string.
\be
\left(
\begin{array}{c}
\eb[1] \\
\eb[2] \\
\tb
\end{array}
\right) \ = \ U \left(
\begin{array}{c}
\nb \\
\bb \\
\tb
\end{array}
\right).
\label{U}
\ee
In the sequel we shall parameterise the rotation matrix $U$ as follows,
\[
U= \left(
\begin{array}{ccc}
\cos\eta(s) & -\sin\eta(s) & 0 \\
\sin\eta(s) & \cos\eta(s) & 0 \\
0 & 0 & 1
\end{array}
\right).
\]
In the case of a structureless string, the frame rotation~(\ref{U}) has no effect on the extrinsic string geometry. But it induces a \emph{gauge} transformation of the Frenet matrix, whereupon the curvature $\kappa(s)$ and torsion $\tau(s)$ become effectively \emph{gauge dependent} quantities, which are subject to the following $\text{SO(2)}\sim\text{U(1)}$ transformation law
\be
\label{gaugesystem}
\begin{matrix}
\kappa \ &\to & \hk & = & {\rm e}^{i\eta(s)} \kappa\,,
\\ \tau \ &\to & \ \htau & = & \tau - \eta'(s)\,.
\end{matrix}
\ee
Here we have extended the Frenet curvature into a complex valued quantity, for a compact presentation of  the ensuing  entries in the rotated Frenet matrix. The transformed $\hk$ and $\htau$ depend on the rotation angle $\eta(s)$ in a manner which is analogous to the effect of a U(1) gauge transformation on the abelian Higgs multiplet, when we interpret curvature as the complex Higgs scalar while torsion is akin the U(1) gauge field. This analogy has been previously employed in~\cite{Danielsson:2009qm,oma-old} to argue that the abelian Higgs model Hamiltonian can be taken as an effective energy function for string-like structures in $\mathbb R^3$, in a manner that generalises the extrinsic curvature based string action proposed in~\cite{poly}, in the context of relativistic strings.

We now proceed to investigate how to extend the proposal of~\cite{Danielsson:2009qm,oma-old}, to construct the most general gauge invariant Hamiltonian governing the dy\-na\-mics of strings in $\mathbb R^3$. We start by introducing a SO(3) valued matrix $\mathcal O(s)$ that we utilise to rotate a given initial frame which is located at an initial point with $s=0$, to a generic point $s$ along the string. This transport matrix $\mathcal O(s)$ acts on the initial frame by the left multiplication, which gives rise to the generalized Frenet matrix
\[
R_F\ \to \ \hat{R}_F \ = \ -\, \mathcal O d_s \mathcal O^{-1}\,.
\]
The $\text{SO(2)}\sim\text{U(1)}$ frame rotations $U(s)$ around the tangent vector $\tb(s)$ act on $\mathcal O(s)$ from the left as follows:
\be
\label{localgauge}
\mathcal O(s)\ \to \ U(s)\cdot \mathcal O(s)\,.
\ee

The dynamics of a structureless string can not depend on the manner how it has been framed. Accordingly, the Hamiltonian must remain invariant under the local transformations~(\ref{localgauge}). For this, we introduce a covariant derivative with a SO(3) valued gauge field $A=A(s)\,T^3$ where $T^3$ is the SO(3) generator of rotations that acts on the local frames in $\mathbb R^3$ by leaving $\tb(s)$ invariant. 
For covariance under frame rotations, the gauge field transforms as follows under $\text{SO(2)}\sim\text{U(1)}$ frame rotations $U$,
$$
A\ \to \ U^{-1} A U - U^{-1}d_sU\,.
$$
We utilise the covariant derivative to introduce the following two gauge invariant quantities
\begin{eqnarray}
L & = & R^{-1}(d_s+A)R\,, \\
L_3 & = & R^{-1}T^3(d_s+A)R\,.
\end{eqnarray}

The most general \emph{quartic} Hamiltonian in terms of $L$ and $L_3$ is then
\begin{eqnarray}
\mathcal H & = & \!\!B_1\tr\!\left(d_sL^Td_sL\right)+B_2\tr\!\left(d_sL_3^Td_sL_3\right) \nn
\\ & + & \!\!B_3\tr\!\left(d_sL_3d_sL_3\right) + B_4\tr\!\left(d_sL_3^Td_sL\right) \nn
\\ & + & \!\!A_1\tr L^TL + C_1\tr\left\{L^T,L\right\}^2 + C_2\tr\left\{L_3^T,L_3\right\}^2 \nn
\\ & + & \!\!K_1\tr L_3+ K_2\tr L_3^2 + K_3\tr L_3^3+ K_4\tr L_3^4 \nn
\\ & + & \!\!F_1\left(\tr L_3\right)\left(\tr L^T L\right).
\label{actionL}
\end{eqnarray}
Here, in the last line, we have added a single multi-trace functional; all the other multi-trace functionals are linearly dependent, at quartic order.

When we proceed and relate the $L$ and $L_3$ with the generalised Frenet matrix, we find the following Hamiltonian in terms of the (generic frame) curvature and torsion fields:
\begin{eqnarray}
\mathcal H & = & b_1(A'-\htau')^2 + b_2(d_s-iA)\hk^\ast (d_s+iA)\hk \nn
\\  & + & b_3 (d_s-i\htau)\hk^\ast (d_s+i\htau)\hk  + b_4(A'-\htau')|\hk|^2 \nn
\\ & +  & a_1 \left|\hk\right|^2 + f(A-\htau)\left|\hk\right|^2 +  c_1(A-\htau)^2|\hk|^2 \nn
\\ & + & c_2|\hk|^4+ \Sigma_{n=1}^4 k_n(A-\htau)^n\,.
\label{actionkt}
\end{eqnarray}
Here the lowercase coefficients are in general linear combinations of the uppercase coefficients in~(\ref{actionL}).

In~(\ref{actionkt}), we identify the following three $\text{SO(2)}\sim\text{U(1)}$ gauge \emph{i.e.} frame rotation invariant combinations,
\begin{eqnarray*}
&&|\hk|^2\,, \\
&&A-\htau\,, \\
&&A+d_s\left(\arg\hk\right).
\end{eqnarray*}
For the gauge degree of freedom $A$, we may proceed, for example, with the identification
\[
A\equiv \htau\,.
\]
This identification has been implicitly used \emph{e.g.} in~\cite{Danielsson:2009qm}. The Hamiltonian~(\ref{actionkt}) then reduces to
\be
\label{actionHiggs}
b_3 (d_s-i\htau)\hk^\ast (d_s+i\htau)\hk   + a_1 \left|\hk\right|^2  + c_2|\hk|^4\,,
\ee
which is precisely the Hamiltonian of the abelian Higgs model. Alternatively, we may choose
\[
A\equiv -d_s\left(\arg\kappa\right)\,.
\]
By redefining $\tau$ as the gauge invariant combination $\htau-A$ and by introducing a gauge invariant variable $\kappa$ via $\kappa^2=|\hk|^2$, we then arrive at the following Hamiltonian
\begin{multline}
\label{actionnew}
b_1(\tau')^2 + \tilde{b}_2(\kappa')^2  - b_4\tau'\kappa^2 + a_1 \kappa^2 + f\tau\kappa^2 \\
+  \tilde{c}_1\tau^2\kappa^2 + c_2\kappa^4+ \sum\limits_{n=1}^4 \tilde{k}_n\tau^n\,.
\end{multline}
Note that for this choice, the gauge field $A$ vanishes in the Frenet frames. Thus, with this choice, the gauge invariant combination $\tau=\htau-A$  coincides with the original, geometric torsion $\tau$ that appears in the Frenet equation; we call this the Frenet gauge. We also point out the consistency of this choice of $A$, with the equations of motion that identify the extrema of the general Hamiltonian~(\ref{actionkt}). The equations yield the following constraint
\be
b_2(A+\eta^\prime)+b_3(\htau+\eta^\prime)=0\,.
\label{constraint}
\ee
This eliminates one of the degrees of freedom, depending on the choice of coefficients $b_2$ and $b_3$. For example, by selecting $b_3=0$ we obtain $A=-\eta'$ and accordingly~(\ref{actionnew}) emerges as a truncation of~(\ref{actionkt}). Moreover, \emph{any} choice with $b_2+b_3\neq 0$ leads to~(\ref{actionnew}) as
a reduced Hamiltonian (with appropriately chosen coefficients).

The alternative $b_2=-b_3$, which implies $A=\htau$, eliminates the derivative (kinetic) term of the curvature. Accordingly it is only consistent with constant curvature strings. Finally, (\ref{actionHiggs}) is a particular case of~(\ref{actionnew}), it corresponds to the choice $b_1=b_4=f=k_n=0$, $\bar{b}_2=\bar{c}_2$ with the identification $\tau=\htau+\eta'$.

We note that constructing a discretised lattice version of the Hamiltonian requires a replacement of the string by a piecewise linear polygonal chain, with the variable $s$ replaced by a discrete index $i=1,\ldots,N$ that labels its vertices. The local curvature and torsion of the string become replaced by the bond and dihedral angles of the chain~\cite{Hu:2011wg}, and at each vertex $s$ the direction of the subsequent polygon section is defined by a rotation matrix $\mathcal R_i$. The discrete analog of the gauge invariant operators $L$ and $L_3$ have the form
$$
L_i \ = \ \mathcal R_{i+1}^{-1}\nabla \mathcal R_i \ \equiv  \ 1- \mathcal R_{i+1}^{-1}T_i \mathcal R_{i}\,,
$$
and
$$
L_i^{(3)}\ = \  \mathcal R_{i+1}^{-1}T^3 \mathcal  R_{i+1} - \mathcal  R_{i+1}^{-1}T^3T_i \mathcal  R_i \,.
$$
Here $T_i$ is a ``parallel transport" matrix which is associated to the link connecting vertices $i\sim s$ and $i+1\sim s+\delta s$; in terms of the gauge field that emerges in the continuum
\[
T(s)=\exp\int_s^{s+\delta s} A(x)\,dx\,.
\]

Again, from the operators $L_i$ and $L^{(3)}_i$, one may construct an invariant Hamiltonian. By discretising~(\ref{actionL}) and replacing integrals by sums we arrive at the following quartic discrete Hamiltonian
\begin{eqnarray}
\mathcal H
\!\!\!\! & = & \!\!\!\!B_1\sum\limits_{i=1}^{N-1}\tr\!\left(L_{i+1}^TL_{i}\right) + B_2\sum\limits_{i=1}^{N-1}\tr\!\left((L_{i+1}
^{(3)})^TL_{i}^{(3)}\right) \nn
\\ & + & \!\!\!\!B_4\sum\limits_{i=1}^{N-1}\tr\!\left((L_{i+1}^{(3)})^TL_{i}\right) + B_3\sum\limits_{i=1}^{N-1}
\tr\!\left(L_{i+1}^{(3)}L_{i}^{(3)}\right) \nn
\\ & + & \!\!\!\!F\sum\limits_{i=1}^{N}\tr L_i^{(3)}\,\tr\!\left\{L_i^T,L_s\right\}^2 +
A_1\sum\limits_{i=1}^{N}\tr L_i^TL_i\nn
\\ & + & \!\!\!\! C_1\sum\limits_{i=1}^{N}\tr\!\left\{L_i^T,L_i\right\}^2
+ C_2\sum\limits_{i=1}^{N}\tr\!\left\{(L_i^{(3)})^T,L_i^{(3)}\right\}^2 \nn
\\ & + & \!\!\!\!\sum\limits_{i=1}^4 K_i \sum\limits_{j=1}^{N}\tr (L_j^{(3)})^n.
\label{discreteaction}
\end{eqnarray}

We now proceed to interpret the gauge field $A$: thus far, this gauge field has been interpreted as an auxiliary construct in the Hamiltonian. Its initial presence ensures gauge invariance, \emph{i.e.} invariance of the Hamiltonian under local frame rotations. But we have found that $A$ can be consistently removed by using the classical equations of motion. Alternatively, we may also continue and retain the gauge field $A$, with an attempt to provide it a physical interpretation. In particular, if we promote $A$ into a full three-dimensional gauge field in the ambient $\mathbb R^3$, we can utilise it to introduce a long range self-interaction between distant segments of a given string, and a long range interaction between different spatially separated individual strings. For this, we start with a bulk gauge field $\Ab(x)$, so that along each string we have for the corresponding local tangent vector
\[
\tb(s) \cdot\Ab(x(s))=A(s)\,.
\]
The Hamiltonian~(\ref{actionL}) then continues to remain invariant under the appropriate $\mathbb R^3$ gauge transformations $\mathcal U(x)$.  Moreover, we may extend the Hamiltonian~(\ref{actionL}), which has support on $r=1,...,N$ separate strings, by adding the three-dimensional Chern-Simons term for $\Ab(x)$
as the gauge invariant interaction term between the separate strings,
\be
\label{CSaction}
\sum\limits_{r=1}^{N}\int ds_r\ \mathcal H_r \ + \ \frac{m}{4\pi} \int d^3x\, \epsilon^{ijk}A_i\partial_j A_k\,.
\ee
Here we assume for simplicity that the individual strings are either closed or infinitely long, to preserve gauge invariance of the Chern-Simons term. Note that additional terms such as in particular the Wilson line integrals $\int A(s)\,ds$ are also gauge invariant, and could be included.

The consistency of the ensuing equations of motion demands that we have
\be
\label{CS2}
\partial_u A_v - \partial_v A_u \ \equiv \ F_\perp  =  \sum_r^N q_r\,\delta^{(2)}(\vec{x}-\vec{x}_r(s))\,,
\ee
where $\vec{x}\equiv(u,v)$ parameterise directions in $\mathbb R^3$ that are orthogonal to the string at the point $s$ (\emph{i.e.} directions in the plane of
the binormal $\bb$ and the normal $\nb$), and $q_r$ are constants. We also obtain the following deformation of the constraint~(\ref{constraint}):
\be
\label{constraint2}
\left(b_2(A+\eta_r')+b_3(\htau_r+\eta_r')\right)|\hk_r|^2\ =\ \frac{mq_r}{\pi}\,.
\ee

We identify here the conventional (abelian) Chern-Simons equations, now described by action~(\ref{CSaction}) and with non-trivial dynamics introduced by the strings in lieu of Wilson lines in $\mathbb R^3$; these strings coincide with the loci of the delta-functions. Accordingly, the Chern-Simons term endows the strings with a topologically determined statistical interaction, at the quantum level. In particular, we conclude that a set of three dimensional closed strings may be subject to fractional statistics~\cite{prl}, which is supported by the presence of the Chern-Simons gauge field $\Ab(x)$.
\vskip 0.3cm

In conclusion, we have constructed the most general quartic Hamiltonian for three-dimensional strings, in terms of their extrinsic curvature and torsion. The starting point in our construction is the concept of frame rotation invariance, with the ensuing rotation matrix of frames as the elemental degree of freedom.
The demand of frame rotation invariance has prompted us to introduce a one dimensional gauge field as an auxiliary degree of freedom, which one can subsequently identify with the torsion variable by using the equations of motion. However, there is also the possibility to consistently promote the gauge field into a field with support in the bulk $\mathbb R^3$, and then endow it with the corresponding Chern-Simons term. This extension introduces a long distance interaction both between spatially distant segments of a given single string, and also between different strings in $\mathbb R^3$. In particular, the presence of a Chern-Simons term proposes that fractional statistics is prevalent, among closed strings in $\mathbb R^3$.

\paragraph{Acknowledgements} This work has been supported by a STINT Initiation Grant between Uppsala University and International Institute of Physics -- UFRN. AN and AS thank the International Institute of Physics, while IG and DM thank the Department of Physics and Astronomy at Uppsala University for hospitality, during the course of the present collaboration. The work of DM constitutes part of the Science without Borders project 400635/2012-7 supported by the Brazilian National Counsel for Scientific and Technological Development (CNPq). The work by DM was also partially supported by Russian RFBR grant 14-02-00627, by cooperation grant 14-01-92691-Ind-a, and by the grant of the support of scientific schools NSh-1500.2014.2. The work of AN has been supported by a grant from Vetenskapsr{\aa}det and by Qian Ren Grant at Beijing Institute of Technology. The work of AS was partially supported by grant 13-1C132 of the Armenian Research Council.

\end{document}